\documentclass[twocolumn,showkeys,preprintnumbers,floatfix,amsmath,a4paper,nofootinbib,amssymb]{revtex4}
\usepackage{graphicx}
\begin{document}
\title{Formation of homophily in academic performance: students prefer to change their friends rather than performance}

\author{Ivan Smirnov$^1$ and Stefan Thurner$^{2,3,4}$}

\thanks{stefan.thurner@meduniwien.ac.at}

\affiliation{
$^1$ Institute of Education; National Research University Higher School of Economics, Myasnitskaya ul., 11, Moscow 101000, Russia\\
$^2$ Section for Science of Complex Systems; Medical University of Vienna, Spitalgasse 23; A-1090, Austria\\
$^3$ Santa Fe Institute; 1399 Hyde Park Road; Santa Fe; NM 87501; USA. \\
$^4$IIASA, Schlossplatz 1, A-2361 Laxenburg; Austria.}

\begin{abstract}
Homophily, the tendency of individuals to associate with others who share similar traits, has been identified as a major driving force in the formation and evolution of social ties. In many cases, it is not clear if homophily is the result of a socialization process, where individuals change their traits according to the dominance of that trait in their local social networks, or if it results from a selection process, in which individuals reshape their social networks so that their traits match those in the new environment. Here we demonstrate the detailed temporal formation of strong homophily in academic achievements of high school and university students. We analyze a unique dataset that contains information about the detailed time evolution of a friendship network of $6,000$ students across $42$ months. Combining the evolving social network data with the time series of the academic performance (GPA) of individual students, we show that academic homophily is a result of selection: students prefer to gradually reorganize their social networks according to their performance levels, rather than adapting their performance to the level of their local group. We find no signs for a pull effect, where a social environment of good performers motivates bad students to improve their performance. We are able to understand the underlying dynamics of grades and networks with a simple model. The lack of a social pull effect in classical educational settings could have important implications for the understanding of the observed persistence of segregation, inequality and social immobility in societies.
\end{abstract}

\maketitle


Homophily is the tendency of humans to associate with others who share similar traits. It has been observed for a 
multitude of different traits, including gender \cite{shrum1988friendship,tuma1979effects}, race 
\cite{currarini2010identifying,shrum1988friendship,tuma1979effects}, academic achievements 
\cite{tuma1979effects,flashman2012academic,lomi2011some}, genotypes \cite{fowler2011correlated}, 
aggression \cite{espelage2003examination}, obesity \cite{christakis2007spread}, happiness \cite{fowler2008dynamic}, 
divorce \cite{mcdermott2013breaking}, smoking \cite{ennett1994contribution}, or sexual orientation \cite{thelwall2009homophily}.
Homophily is found for different types of relationships such as between spouses \cite{blossfeld2009educational}, 
friends \cite{christakis2013social} and co-workers \cite{ibarra1992homophily}, and occurs in a wide range of environments 
including kindergarten \cite{hanish2005exposure}, large human gatherings \cite{barnett2016social},  Wall Street \cite{roth2004social}, 
populations of hunter-gatherers \cite{apicella2012social}, or virtual societies of online gamers \cite{szell2010multirelational,szell2013women}.
Homophily is considered as one of the fundamental organizational principles of human societies \cite{mcpherson2001birds}, 
and has a number of important social implications such as the origin of segregation \cite{currarini2009economic} or  
the perpetuation of economic inequality and social immobility \cite{dimaggio2012network}.

\begin{figure}[th!]
\begin{center}
\includegraphics[width= 5.2cm]{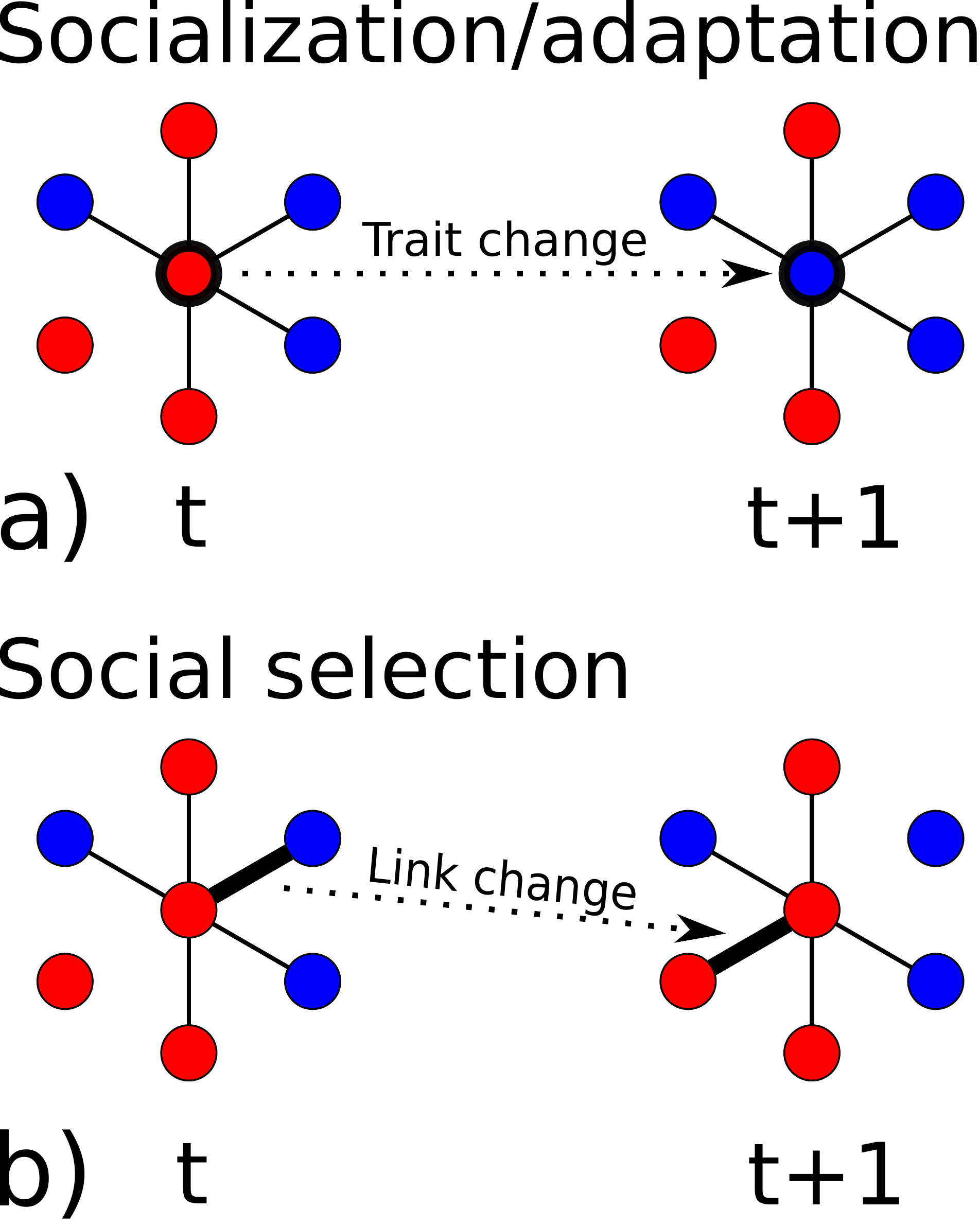}
\caption{Two basic mechanisms to understand the origin of homophily. Nodes represent individuals, different colors indicate different traits. 
Links correspond to social ties, e.g. friendship. Increase of similarity between connected individuals may either arise from changes in traits 
(socialization process)  (a), where individuals change their trait according to the dominance of that trait in their local social networks,
or through re-wiring of their local social networks (social selection process) (b), where individuals re-shape their social contacts such that their trait
matches those in the new environment better.}
\label{Fig:mechanisms}
\end{center}
\end{figure}

Even though there exists an extensive body of research on homophily, it remains a challenging question to understand its origins and how it forms and develops
over time. For traits that can not be changed, such as race or gender, homophily arises through a re-structuring process of inter-human relationships,
where on average links between people with similar traits are created, while links between dissimilar people are dissolved.
If traits can be changed over time, the situation becomes more involved. In this case, there exist two mechanisms to explain the formation of homophily
from an initially homogeneous population: {\em socialization} and {\em social selection} \cite{kandel1978homophily}.
The mechanism of socialization means that people change their traits to increase similarity to those they are connected to in a static social environment (network).
This is schematically shown in Fig. \ref{Fig:mechanisms} (a). This mechanism is sometimes also referred to as `social contagion' or `peer influence'.
Under the mechanism of social selection, individuals re-arrange their social ties so that they become linked to people that are similar in traits, see
Fig. \ref{Fig:mechanisms} (b). If both mechanisms are at work at the same time, traits and social networks are said to {\em co-evolve}. The literature on the aspect of co-evolution is limited because of the lack of the simultaneous availability of longitudinal social networks and traits data.
Homophily in academic performance has been studied with network data collected with questionnaire-based surveys \cite{flashman2012academic,lomi2011some}.
The design of these studies makes it hard to follow the temporal evolution of social networks.
The availability of new technologies and big datasets provides researchers with novel tools to observe the dynamics of social networks with high temporal precision.
For example, the temporal structure of social networks has been reconstructed by email data \cite{kossinets2006empirical}, computer game logs\cite{szell2010measuring}, or interactions on learning management system platforms \cite{vaquero2013rich}.

\begin{figure}[t!]
\begin{center}
\includegraphics[width= 8cm]{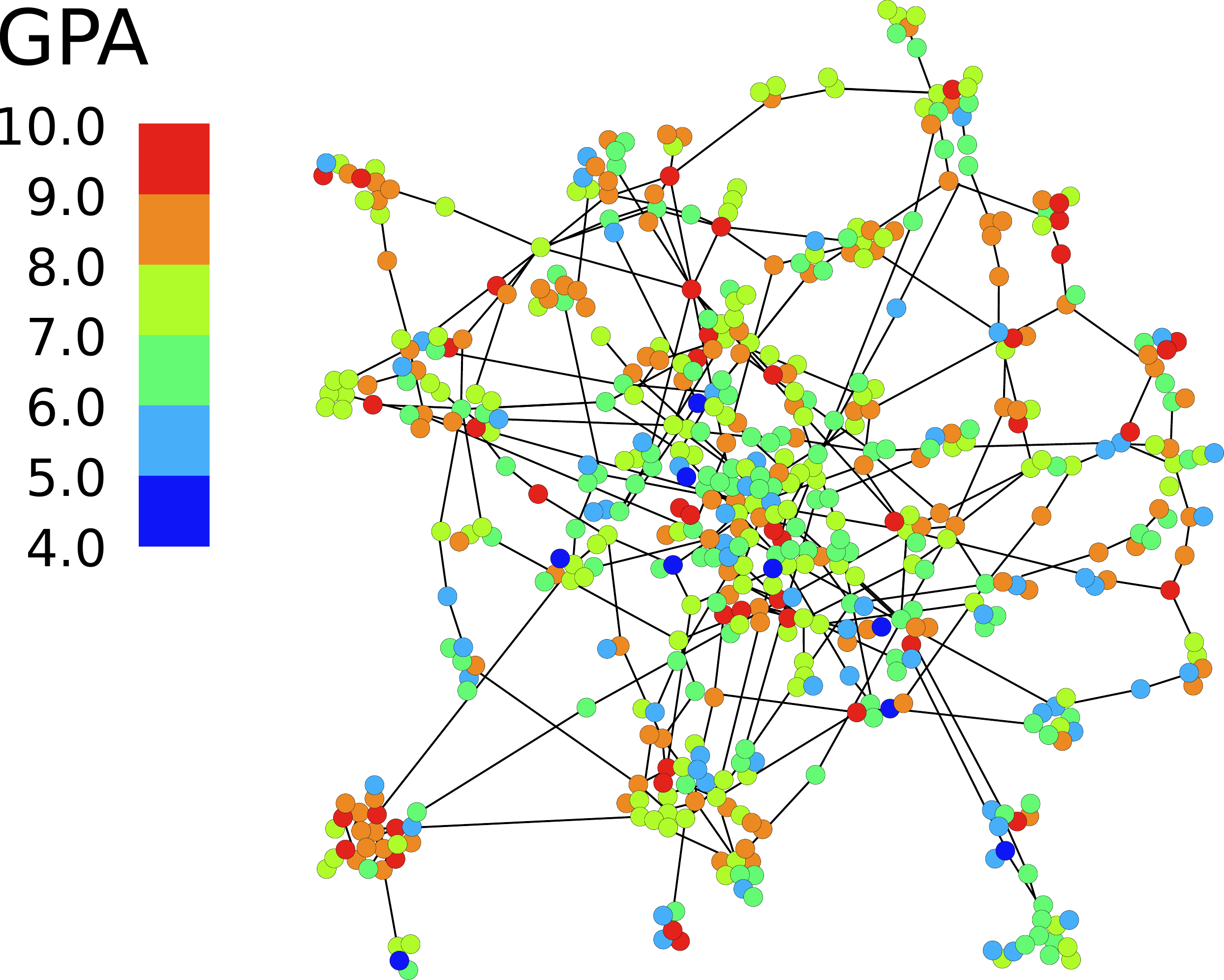}
\caption{Snapshot of the friendship network of university students. The network is reconstructed from students' interactions on the social network site VK, the Russian variant of Facebook. Nodes represent students, links exist if one student gave a "like" to another at least once in March 2016.
Color represent the performance (GPA) of students across the whole period of studies. There is visible clustering of students with similar GPA.}
\label{Fig:network}
\end{center}
\end{figure}

In this paper, we use a unique anonymized dataset to observe the temporal formation of academic homophily based on social interactions between
Russian students from a public high school and a university. The dataset contains information on the students' academic 
performance at several time points during their studies together with the detailed information about the evolution of their friendship networks (see SI).
These networks between students were obtained from the largest European social network site VK\footnote{http://vk.com}, 
that provides a functionality similar to Facebook. VK users create their profiles with information about their identity, education, interests, etc.
The use of the real name is required by VK. Users may indicate other users as their friends. VK friendship is mutual and 
requires confirmation. However, using VK friendship links is not the most efficient way to study the
dynamical evolution of actual friendships, since only information about the current friendship is available, which  
makes it practically impossible to extract the dynamics of VK friendship links. It is also impossible to distinguish active 
friendship ties from obsolete ones since VK friendship links are rarely dissolved. 
We, therefore, approximate friendship links between students from their activities on the social networks site, in particular from the placement of
``likes'' \cite{scissors2016s} on other students' pages. A link from one student to another is created if a ``like'' was placed at least once within 
a given period of time (see SI). This approximation of actual friendships by social interaction strengths 
allows us to track the effective network evolution between students with much higher precision (see SI Fig. \ref{Fig:history}).
The network of university students (seniors) on March 2016 is shown in Fig. \ref{Fig:network}.
Previous studies on the Facebook (or its Russian analog VK) have focused on the (relatively static) friendship marking options that 
are provided by the sites \cite{mayer2008old,lewis2012social,dokuka2015formation}.
This dataset not only allows us to quantify the extent of academic homophily among students but also to see its detailed evolution over time.
In particular, we are able to clarify the mechanism behind the emergence of academic homophily from an initially homogeneous population 
across several years.

We use two datasets of academic performance records measured as grade point averages (GPA), one with 655 students from the 5th to 11th 
grades (age from 11 to 18) of a Russian public high school in Moscow\footnote{For reasons of anonymity we do not state the name of the school}, 
the other with 5,925 bachelor students of the Higher School of Economics in Moscow. 
High school students receive their grades at the end of each trimester, their GPAs for the last 5 trimesters were available. Since the academic year of 2014/15  the Higher School of Economics started to publish a public ranking of its students. It contains information about their GPAs for the current semester along with the aggregated average GPA across the whole period of their studies. We collect the temporal  GPA data into a vector $G^{\rm HS/U}_{i}(t)$ that represents the GPA of student $i$ at time $t$ and corresponds to a student's performance within the time period from $t-1$ to time $t$. $t=1$ indicates the end of the first trimester/semester, $t=T$ is the end of the last trimester/semester for a given group of students. ${\rm HS}$ indicates ``high school'', ${\rm U}$ is ``university''. For detailed information about time points corresponding to GPAs collection see SI Fig. \ref{Fig:data}. Note that grades are different for high school and university. For high school grades range from $2$ (worst) to $5$ (best), for university from $4$ (worst) to $10$ (best). The average GPA of a student across the entire available time period we denote by $\bar G^{\rm HS/U}_i$. For university students we follow $4$ cohorts that are labeled by $X$ in the following way: 
$G^{\rm U, X}_{i}$, where  $X = 1,2,3,4$ stands for freshmen, sophomores, juniors, and seniors, respectively. The average GPA for high 
school students and the cohorts of university students are presented in SI Table \ref{Tbl:gpa_statistics}.

To generate a proxy for the temporal friendship interaction network between students we use the popular SNS VK, whose main  
component is a user-generated news feed. This feed contains all content that was generated (posted) by users and is generally visible to friends only. 
If users like the  content that was posted by their friends they can indicate this by an instant feedback called a ``like''. ``Likes'' may mean 
different things to different people \cite{scissors2016s}, however, ``likes'' can, in general, be seen as an indication of {\em active} friendship contacts between users.

VK provides an application public interface (API) that allows to download information systematically in an open JSON\footnote{http://json.org} format. 
In particular, it is possible to download user profiles from particular educational institutions and within selected age ranges. 
For each user, it is possible to obtain the list of their friends and the content that was published by them along with the VK identifiers of users 
that liked this content. Posting times are known with a time resolution of one second.  
``Likes'' for specific content are almost always placed within 1-2 days after the content was posted.
Using specially developed software the profiles of students of a given institution were downloaded and automatically matched by their first and last names with the available data on students' performance. 88\% of all high school students and 95\% of university students could be identified on VK (see SI).
The matching procedure was performed by authorized representatives of the high school and the Higher School of Economics, respectively. 
After the matching procedure, all names and VK identifiers were irrevocably deleted. 
The ``likes'' of all users were collected with corresponding timestamps, those from users 
outside the educational institutions were removed. ``Likes'' were then aggregated to intervals of $3$ months periods. For each group of students, we obtain a $N\times N$ adjacency matrix $A(t)$, where $A_{ij}(t) = 1$ if student $i$ places at least one ``like'' to student $j$ from time $t-1$ to time $t$. For detailed information about time periods corresponding to collected network data see SI Fig. \ref{Fig:data}.
The subsequent deletion of all information on individual ``likes'' and respective timestamps prevents the possibility of any de-anonymization. The resulting datasets were transferred to the Institute of Education, which made it available for research in fully anonymized form.


\section*{Results}
\begin{figure}[t]
\begin{center}
\includegraphics[width= 7.5cm]{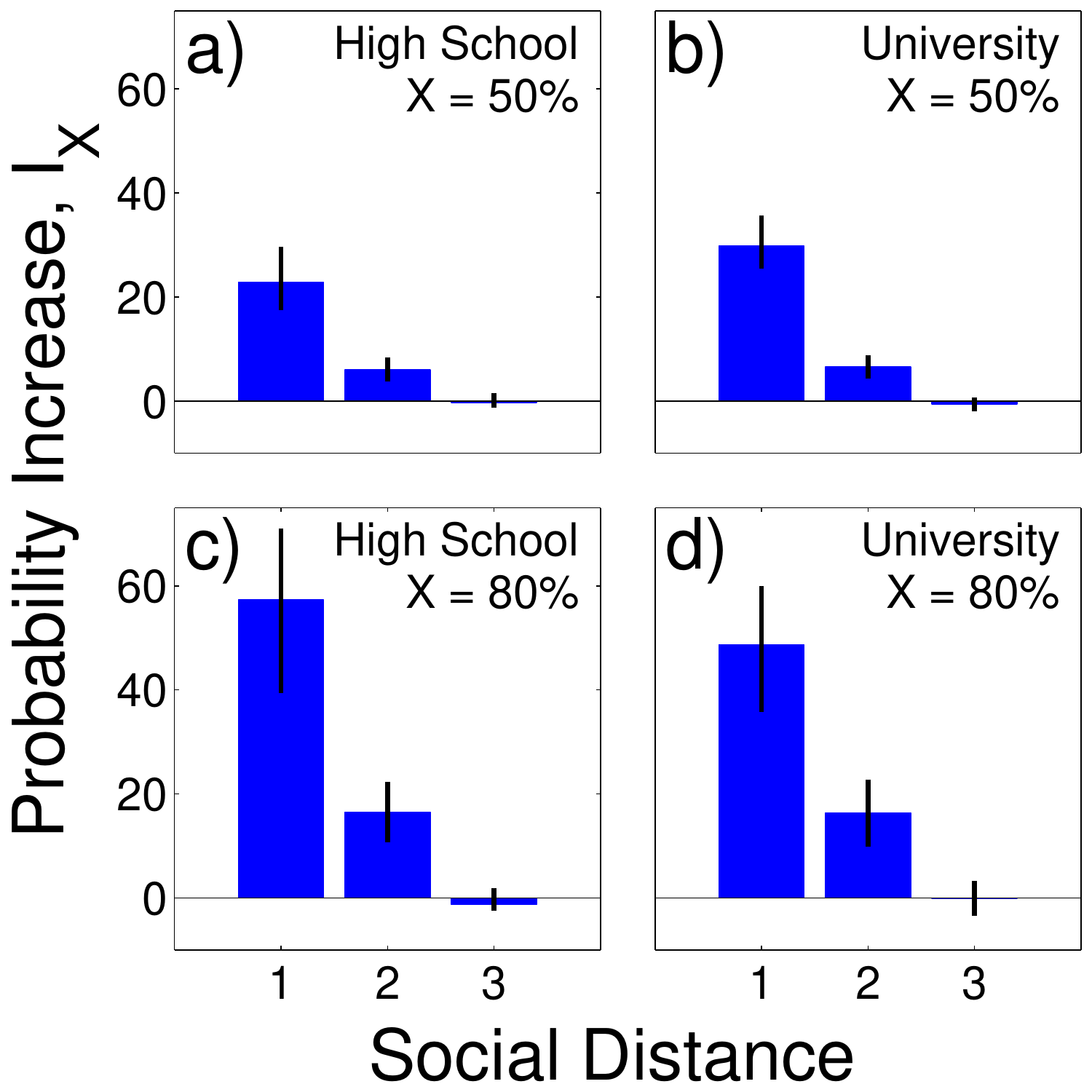}
\caption{Homophily of students with good (a) and (b), and excellent grades (c) and (d), as a function of social distance. 
Observed increase in probability $I_X$ that a student is in the top $X$th percentile of 
students, given that their friends are also in the top $X$th percentile. 
Results for the  high school are shown in  (a) and (c), for university in (b) and (d). 
Vertical lines indicate 95\% confidence intervals computed with the permutation test. 
The social distance of 1 means friends, the social distance of 2 means friends of friends and the social distance of 3 means friends of friends of friends.}
\label{Fig:probability_increase}
\end{center}
\end{figure}

We first demonstrate the existence of academic homophily and then try to understand its origin.  
For all groups of students we find strong homophily. To make it comparable with other homophily studies such as in \cite{christakis2013social}, we 
use a standard way of quantifying it by the conditional probability increase, $I_X$ that a student belongs to top $X$th percentile of performers, given that his/her friends also belong to the same percentile (see SI). 
$I_{X}(t) = 0$ means that grades and friendship network are uncorrelated, $I_{X}(t) = 100\%$  means that the probability to 
be in the top $X$th percentile is 2 times higher if the student's friends are also in the top $X$th percentile, compared to the situation when they are not.
$I_{X}(t)$ can not only be computed for friends (social distance 1) but also for friends of friends (social distance 2), and friends of friends of friends (social distance 3), etc. 
In Fig. \ref{Fig:probability_increase}  we fix $X$ to be the 50th (above average students) (a) and (b) and 80th (excellent students) percentile (c) and (d), 
respectively. For social distances up to 2 we observe significant homophily for all student groups at the last time point $T=6$ for high school, 
and $T=14$ for university. We find 
$I^{\rm HS}_{  50\%}(6)   = 23\%$, 
$I^{\rm HS}_{  80\%}(6)  =  57\%$, 
$I^{\rm U, 4}_{50\%}(14) = 30\%$, 
$I^{\rm U, 4}_{80\%}(14) = 49\%$, $p$-value $< 10^{-4}$. 
Significance was tested with a permutation test (10,000 permutations), see Methods. 
Note that the corresponding values at the first time point are smaller,  
$I^{\rm HS}_{  50\%}(1) = 22\%$, 
$I^{\rm HS}_{  80\%}(1) = 34\%$, 
$I^{\rm U, 4}_{50\%}(1) = 16\%$, 
$I^{\rm U, 4}_{80\%}(1) = 28\%$.

This result holds independent from the method used. Following an alternative approach for scalar variables we compute the {\em assortativity coefficient} $r$ 
\cite{newman2010homophily} and again find highly significant homophily at the last time point 
$r^{\rm HS}(6) 	 = 0.20$ ($p$-value $< 10^{-4}$)  and 
$r^{\rm U,4}(14) = 0.21$ ($p$-value $< 10^{-4}$). At the first time point homophily is much smaller, 
$r^{\rm HS}(1) 	 = 0.12$ and 
$r^{\rm U,4}(1)  = 0.12$. 

\begin{figure}[t]
\begin{center}
\includegraphics[width= 4.2cm]{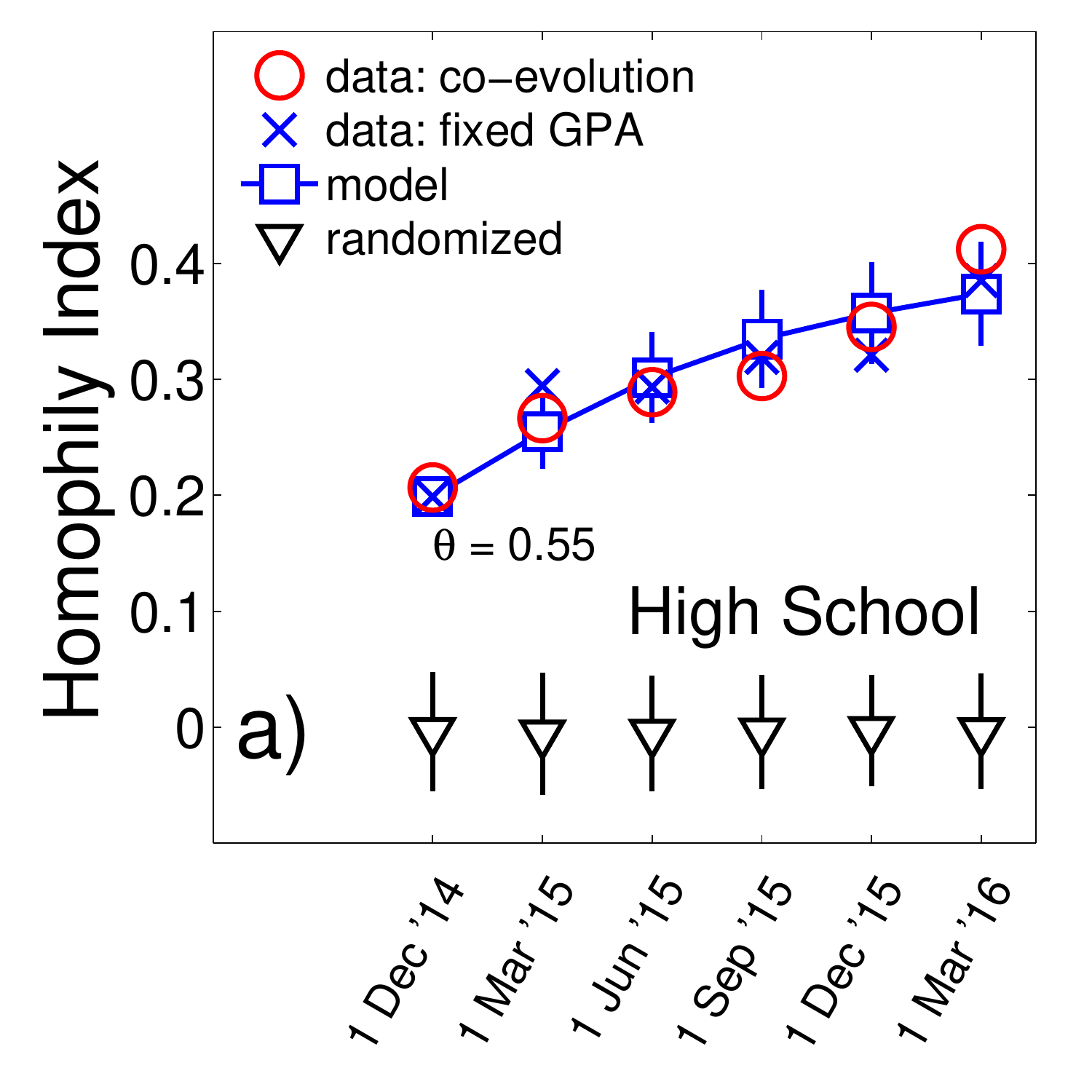}
\includegraphics[width=4.2cm]{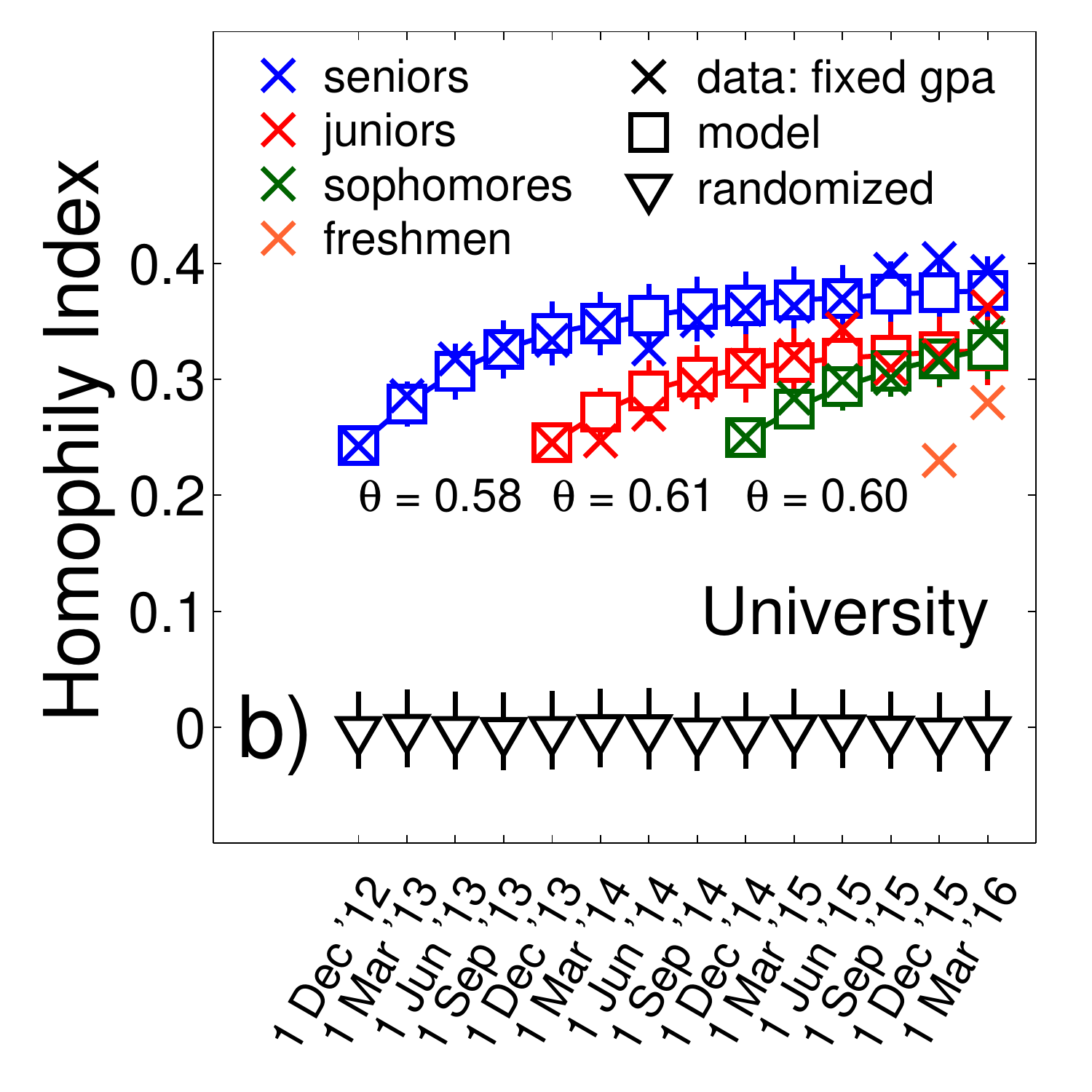}
\caption{Evolution of homophily (Homophily Index) in friendship networks of high school (a) and university students (b). 
Homophily increases with time by almost a factor of 2 (circles).  
The significance of the observed effect is measured with a randomization test (triangles), where grades were 
reshuffled randomly between the nodes in the network. 
It is amazing that when the GPAs of individual students are fixed to their temporal average (crosses), practically the same 
increase of homophily is observed, which signals the dominance of network restructuring.
Results can be understood with a simple model (squares). Vertical bars are standard deviations.
}
\label{Fig:results}
\end{center}
\end{figure}

In Fig. \ref{Fig:results} we show the time evolution of homophily over 1.5 years for high school students (a) and over 3.5 years for university students (b).  
We employ a transparent definition of a {\em Homophily Index}, $H$(see Methods).
We see a clear increase of $H$ from the first  to the  last trimester from about $H=0.20$ to $H=0.41$ for the high school (a) (circles), 
and from  $H=0.24$ to $H=0.40$ for university (b) (crosses). 
We next show in a series of three arguments that the increase in homophily over time can not be explained by the 
socialization/adaptation mechanism, i.e. by the changes in GPAs over time. 

\subsection{Ruling out socialization/adaptation}
(i) The first argument why  socialization/adaptation is not the relevant mechanism behind the observed 
homophily increase, is due to the fact that academic performance is known to be a relatively persistent feature of students. 
It was shown that school-entry academic skills have large predictive power for later academic performance \cite{duncan2007school}, and that academic performance might be heritable \cite{krapohl2014high}.
We find the persistence of performance in our data. The average GPA over high school students ($3.85\pm0.55$) and its variance 
do practically not change over time, see SI Fig. \ref{Fig:grade_diff}. 
The average absolute difference between two consecutive time points $\langle |G_i(t) - G_i(t-1)| \rangle_{i}$ is $0.130$, 
which means that the variation between GPAs of the same student at different time points is much smaller 
than the variation across students. Similar results are observed for the university students, with an average GPA of $7.41 \pm1.03$ 
and a mean absolute difference $0.40$.

(ii) The second argument  why the socialization/adaptation mechanism can be ruled out is due to the observation that if we fix the GPAs 
for the high school students and do not let evolve them over time (we use the average GPA over all trimesters $\bar G_i$), we observe 
practically the same homophily increase as for the co-evolving GPAs, Fig. \ref{Fig:results} (a). 

(iii) Finally, we use a regression model to explain the GPA of students $G_i(t)$ by the explanatory variables: GPA at the previous trimester/semester $G_i(t-1)$, 
by the influence of friends' GPAs, by gender and by age (see SI). The results are presented in SI Table \ref{Tbl:regression}. 
For high school and university alike we find that the coefficients for $G_i(t-1)$ ($\alpha_1$) and gender ($\gamma$) are significant and the coefficient for friends' GPA ($\alpha_2$) is not. Again, this suggests that GPAs are rather stable over time and are almost fully determined by the GPA at the previous time point. 
The regression shows no evidence for an adaptation effect. 

\subsection{Social selection and network re-organization}
Due to the second argument above the explanation of the observed homophily increase can only come 
through changes in social networks over time, i.e. the social selection mechanism, where students preferentially 
select new friends that are similar in performance. 
A simple model allows us to understand the situation. 
It assumes that whenever students select new friends they prefer students who are more similar to them than their current friends. 
Every student $i$ is endowed with a fixed GPA $\bar G_{i}$ (constant).  There exists an initial friendship network that we initialize with 
the observed network at timestep 1,  ${A}^{\rm model}_{ij}(1)=A_{ij}(1)$. From time $t$ to $t+1$ the model runs through the following steps 
\begin{itemize}
\item Pick a student $i$ at random,
\item Pick a random friend $j$ of $i$,  ($A^{\rm model}_{ij}(t) = 1$),
\item Pick a random potential new friend $k$ ($A^{\rm model}_{ik}(t) = 0$),
\item If $k$ is closer to $i$ than $j$, i.e. if $|G_i - G_k| \le |G_i - G_j|$, rewire the link from $ij$ to $ik$. Otherwise, 
rewire the link from $ij$ to $ik$ anyhow, with probability $\theta$,
\item Repeat until all students are updated, then continue with next timestep until $t = T$.
\end{itemize}
Clearly, if $\theta = 0$, rewiring happens only if a potential friend is closer in GPA than a current one (strict homophily increase); 
if $\theta = 1$ we have pure random rewiring. 
For a fixed $\theta$ we compute the Homophily Index ${H}^{\rm model}(t)$ based on model networks. 
$\theta$ is fitted from the data such that $\sum_{t=1}^T ({H}^{\rm model}(t) - H(t))^2$ is minimized. 

We find $\theta$ values within the range of $0.55$ and $0.61$ for all groups. 
This means that students choose new friends among those who are similar about 64\%-81\% more often than among those who are not similar.
The results of the model are presented in Fig. \ref{Fig:results} (boxes). The experimental homophily increase is recovered. 
Remarkably, for all student groups, the model is able to reproduce even details in the empirical GPA distances 
between stable, discontinued, and new friendships, see SI Table \ref{Tbl:reorganization}.  

\begin{figure}[t]
\begin{center}
\includegraphics[width= 5.0cm]{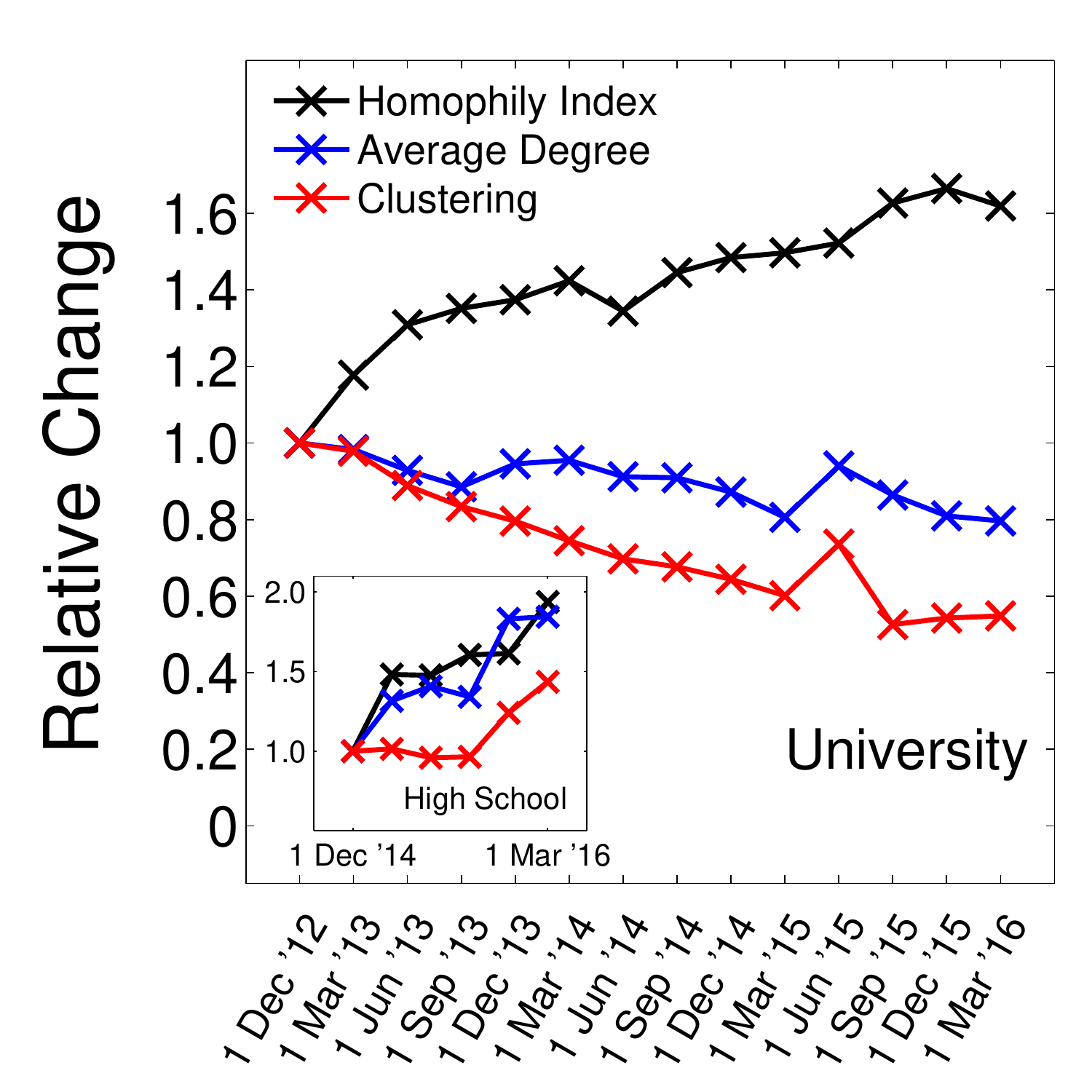}
\caption{The network properties degree and clustering change over time (relative changes are shown, first time point is 1). 
While the network of seniors becomes sparser, there is a densification of the high school network (inset). 
Therefore degree and clustering coefficients can not be the drivers behind the observed homophily increase in 
both groups.}
\label{Fig:network_prop}
\end{center}
\end{figure}

We have to show that the homophily increase is not explained as a trivial consequence of network densification. 
In both datasets we observe that  friendship networks are dynamically changing over time. In Fig. \ref{Fig:network_prop} 
the relative change of the average degree and the clustering coefficient of the networks are shown in comparison with the relative change of homophily.
To see that the observed homophily increase is not a trivial consequence of network densification, observe that while for the high school degree and 
clustering increase, for university (seniors) they decrease. In both cases homophily increases. This is a first indication that degree and clustering are not the drivers of homophily change.
As a second indication we test if $H$ and $I_X$ are significant with respect to a permutation test that preserves network topology. This is indeed the case  
(see Methods). Thirdly, by re-defining time intervals in a way that for each time interval the average degree is approximately the same, we find the same homophily 
increase (see SI Fig. \ref{Fig:network_adj}), indicating that the degree is not an explanatory variable.

Finally, in SI Fig. \ref{Fig:gender} we show that there exist slight gender differences in the homophily increase. 
While both genders show about the same increase over time, the homophily index $H$
is slightly larger for females in the sophomore and senior groups, and larger for males for the high school students and juniors. 

\section*{Discussion}
We studied a unique dataset containing the academic performance of high school and university students 
together with detailed information about the evolution of their social ties. In accordance with previous research \cite{tuma1979effects,flashman2012academic,lomi2011some,mayer2008old} we found strong homophily in academic 
performance. 
The strength of academic homophily is found to be stronger than for homophily in sexual activity \cite{brakefield2014same} 
or alcohol abuse among adolescents \cite{shakya2012parental} but weaker than for homophily in smoking marijuana 
\cite{shakya2012parental}, or for age \cite{newman2010homophily}. 

We are not only able to demonstrate the strong homophily in academic performance but also to monitor  
how it emerges from a homogenous population and how it solidifies over time.  
We show that the observed gradual homophily increase can be explained predominantly by the process of social selection, 
meaning that students re-arrange their local social networks to form ties and clusters of individuals that have similar performance 
levels. 
We could exclude the alternative explanations of social adaptation and co-evolution of social ties and performance.  
With a series of tests we ruled out the possibility that the increase of homophily results from adapting their academic 
performance to the one by their close friends. As an important consequence, this means that there are no indications 
for a pull effect, where groups of friends with good grades stimulate poor performing friends to increase their performance. 
The opposite effect of a negative group influence on students is also not found. 
It can be concluded that academic homophily in the studied groups arises and strengthens almost entirely through network re-linking. 

We are able to understand the social-selection based homophily increase with a simple dynamical one-parameter model. 
The estimate of the parameter from the data means that students choose a new friend among those who are similar to them 
64\%-81\% more often than dissimilar ones. 

Note that even though this model is much simpler than others previously used \cite{snijders2010introduction}, remarkably it is able to 
recover the increase over the whole time period for all groups, and even allows to understand details of the dynamics. It would be interesting  to see in further work if these findings hold more generally also true for other student groups with different social contexts 
and in different countries. 

Our findings might shed light or even confirm that access does not necessarily lead to equity. 
We find indications that physical mixing of students in the same educational institution does not lead to a homogeneous mixing of social ties. 
Even if the initial distribution is rather homogenous, students constantly re-organize their social network during the studies, which 
eventually results in segregation by academic performance. It is possible to conjecture that this mechanism is potentially 
reinforced by the accessibility of modern information technologies where maintaining links does not require physical presence anymore.
Social networks play a crucial role in social mobility \cite{granovetter1995getting,podolny1997resources}. 
A selective university may provide a unique opportunity to create ties that will benefit students in the future. 
However, if initially low-performing students from a disadvantaged background predominantly create ties with 
other lower-performing students it significantly reduces the possibility of upward social mobility and may explain 
the persistence of inequality in societies.


\section*{Methods}
\subsection*{Homophily Index}
We introduce a Homophily Index, $H$, as the Pearson correlation coefficient between the vector of 
students' GPAs, $G_i(t)$, and the vector of the average of the GPAs of their direct friends, 
\begin{equation}
H(t) = \mathrm{corr} \left( G_i(t), \frac{\sum_j A_{ij}(t)G_j(t)}{\sum_j A_{ij}(t)} \right) \quad .  
\end{equation}
If students' grades are independent from average grades of their friends than $H(t) = 0$. 
Positive $H(t)$ means that better average grades of friends lead to better average grades of students and negative 
$H(t)$ means that better average grades of friends lead to worse students' performance. 
$H(t) = 1$ means a linear relation between students' performance and average performance of their friends.

\subsection*{Randomization test}
One of the challenges in understanding correlations of traits between connected individuals is to test  
if the observed homophily effect is significant or if it results trivially from the topology of the underlying network. 
To test for this we employ a typical permutation test, see e.g. \cite{christakis2013social}, where 
we preserve the network topology and randomly reshuffle the assignment of the GPAs to the node. 
We repeat this procedure $10,000$ times to obtain a distribution of the measures $H$ and $I_X$. 
We can then test the null hypothesis that GPAs are independent of network topology, and to compute corresponding 
$p$-values. 

\subsection*{Acknowledgment}
We thank Dr. Alexander Sidorkin from the Institute of Education, Higher School of Economics, Moscow, 
for providing us with the dataset. 
I.S. is grateful to the {\em Advanced Doctoral Program Grant} of the Higher School of Economics, Moscow, Russia.

\bibliography{homophily}

\clearpage

\makeatletter 
\setcounter{figure}{0}
\renewcommand{\thefigure}{S\@arabic\c@figure}
\renewcommand{\thetable}{S\@arabic\c@table}
\makeatother


\section*{Supporting Information (SI)}
\subsection*{SI Text}
\subsubsection*{Academic performance records of high school students}
Children start their compulsory education at the age of 6-8 in Russia. The first 9 years of studies are compulsory. A significant number of students continue their education for two more years. Students study together in the same fixed groups for many years. There is no ability grouping. Group work is generally not practiced. 

Our dataset contains academic performance records of 655 students of a Russian public school. Records include the information about all students from the 5th to 11th grades. From the 5th to 9th grades the average size of the cohort is 108, 44\% of students are girls and 56\% are boys. For the last two grades the average size of the cohort is smaller and the gender balance is reversed. The average cohort size is 56, 56\% are girls, 44\% are boys.

At the end of each trimester, students receive grades for each school subject. Grade can be 2 (not passed), 3 (passed), 4 (good) and 5 (excellent). To assess the average performance GPA is computed for 8 subjects which are present for all cohorts of students (mathematics, physics, informatics, biology, Russian, English, literature, history).

The data $G^{\rm HS}_i(t)$ was collected for the $3$ trimesters of the academic year of 2014/15 and for the first $2$ trimesters of the academic year of 2015/16 (see Fig. \ref{Fig:data} (b)). As students do not study in summer, we assume the same performance as at the last available time point i.e. spring. That is indicated by the equal sign in Fig. \ref{Fig:data} (b). $\bar{G}^{\rm HS}_i$ is computed as the average over the $5$ trimesters. Note that when we compute average we do not count spring grades twice. The average GPA for all students and a comparison between males and females are presented in Table \ref{Tbl:gpa_statistics}.

\subsubsection*{Academic performance records of university students}
The Higher School of Economics in Moscow ranks its students according to their performance. For that purpose, the GPAs of each student is computed. The composition of courses that are used to compute GPA varies from student to student as they are free to choose different courses. The university assigns different weights to different courses to produce the resulting GPA and we use this value ``as is''.

\begin{figure}
\begin{center}
\includegraphics[width=17.8cm]{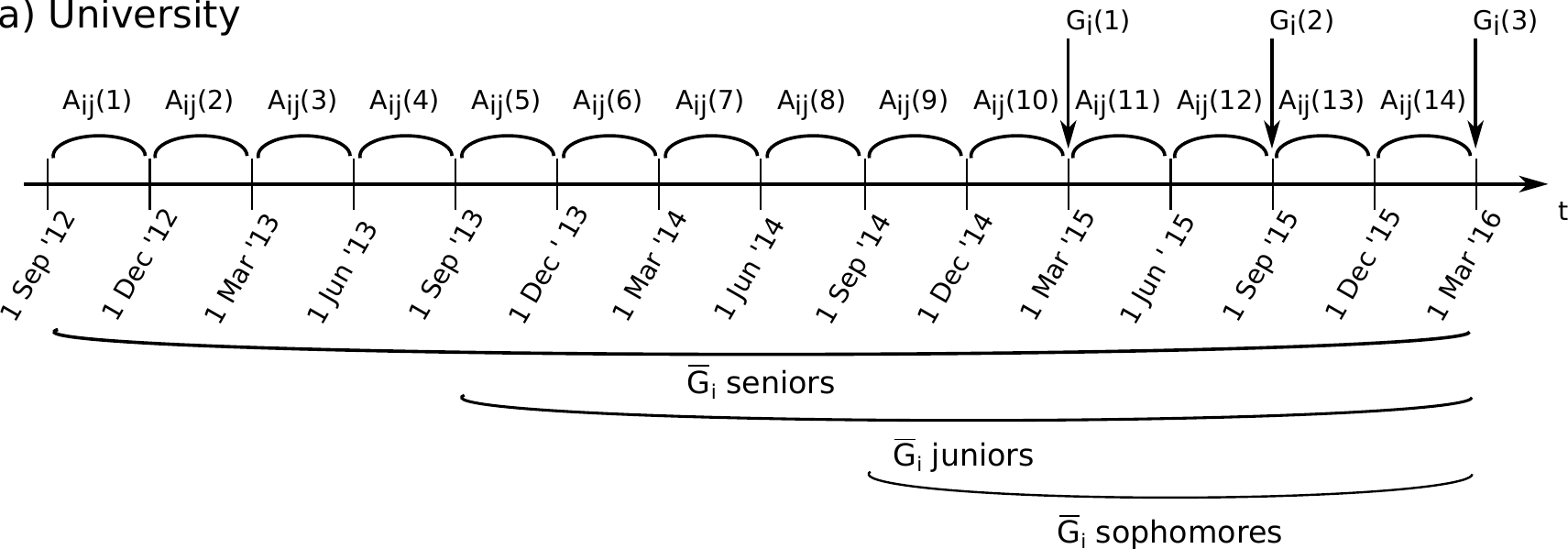}
\includegraphics[width=8.8cm]{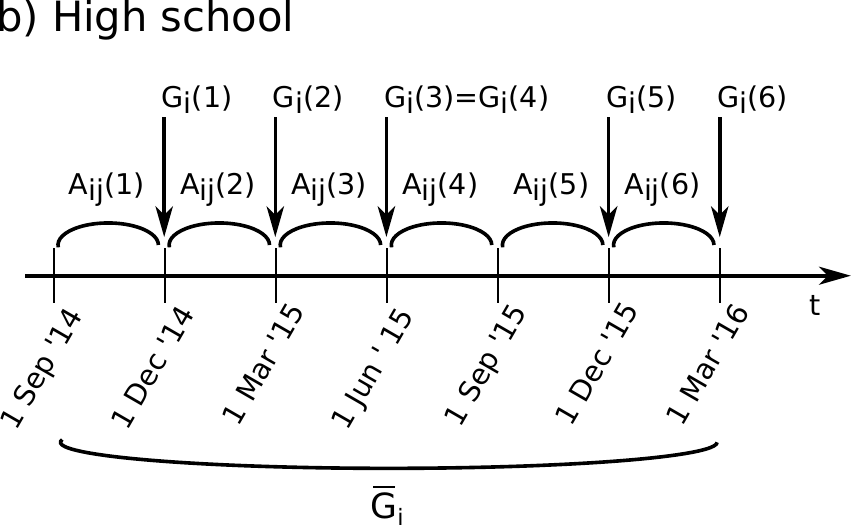}
\caption{
Time schedule of data collection for university (a) and high school (b) students. Network data is in the form of adjacency matrices $A_{ij}(t)$, where $A_{ij}(t) = 1$ means that student $i$ gave at least one ``like'' to student $j$ from time $t-1$ to time $t$. The time period from $t-1$ to $t$ is equal to $3$ months. 
(a) For the university students (seniors, juniors, sophomores) the aggregated average GPA, $\bar{G}^{\rm U}_i$, from the beginning of their studies on the $1$st of September (2012/2013/2014) until the $1$st of March, 2016 is collected. This period is equal to 3.5 years for seniors, 2.5 years for juniors and 1.5 years for sophomores respectively. The temporal GPA data, $G^{\rm U}_i(t)$, was also collected for the last $3$ semesters for all $3$ cohorts (arrows).
(b) For the high school students the temporal GPA data, $G^{HS}_i(t)$, is collected at the end of each trimester for the last $5$ trimesters (arrows). As students do not study in summer, we assume the same performance at that period as at the last available time point i.e. spring, $G^{HS}_i(3) = G^{HS}_i(4)$. $\bar{G}^{HS}_i$ is computed as the average over the $5$ trimesters.
}
\label{Fig:data}
\end{center}
\end{figure}

The university started to publish the ranking of its students openly on its website since the academic year 2014/15. It publishes the academic performance for the current semester and the average GPA, aggregated from the beginning of studies to the present day. In March 2016, these cumulative GPAs for the whole period of studies $\bar{G}^{\rm U}_{i}$ were collected. As indicated in Fig. \ref{Fig:data} (a), this period is equal to 3.5 years for seniors, 2.5 years for juniors and 1.5 years for sophomores respectively. Our dataset also includes GPAs for individual semesters $G^{\rm U}_{i}(t)$, $t = 1,...,3$. $t = 1, 2$ corresponds to the first and second semesters of the academic year of 2014/15 and $t = 3$ corresponds to the first semester of the academic year of 2015/16 (see Fig. \ref{Fig:data} (a)). For freshmen GPAs are available at a single time point corresponding to the first (and only) semester of their studies. The grades are ranging from 4 (worst) to 10 (best). The data contains information about 1,579 freshmen (49\% females), 1,570 sophomores (56\% females), 1,539 juniors (52\% females) and 1,237 seniors (53\% females). The average GPA for each cohort is presented in Table \ref{Tbl:gpa_statistics}.

\subsubsection*{VK data and sampling bias}
VK is the largest European social network site with more than 100 million active users. It was launched in September
2006 in Russia and provides a functionality similar to Facebook. 

According to the VK Terms of Service: ``Publishing any content on his / her own personal page including personal information the User understands and accepts that this information may be available to other Internet users taking into account the architecture and functionality of the Site''. 

There exists the following sampling bias in the data downloaded from VK. The students that  were not identified on the VK are more likely to be males and tend to have lower scores. The proportion of boys among high school students that were not identified is 58\%, and among identified students is 54\%. The proportion of males among not identified university students is 75\% and among identified students is 47\%. The GPA for not identified high school students is 3.77, 3.85 for identified and for university it is 7.00 for not identified and 7.20 for identified students. The proportion of males is significantly higher for not identified university students. However, this bias should not be relevant for the results, as these populations represent less than 5\% of the total population.

\subsubsection*{Homophily measures}
In addition to the Homophily Index as defined in the main text we use two standard ways to quantify homophily.

\begin{enumerate}
\item
The conditional increase in probability is one of the standard ways to quantify homophily for binary variables and was used to demonstrate homophily in obesity, smoking, sleep, heavy drinking, alcohol abstention, marijuana, happiness, loneliness, depression, smiling in profile picture, divorce \cite{christakis2013social}. In the same spirit, from the GPA data $G_{i}(t)$ we can define binary variables, $G^{\rm bin}_{X, i}(t)=1$, if student $i$ ranks above the $X$th  percentile of GPA, and $G^{\rm bin}_{X, i}(t)=0$, if the student ranks below.  

We define $P_{X}^{+}(t)$ as the conditional probability that a student is above the $X$th percentile of GPA given that his or her friend is also above the $X$th percentile, 
\begin{equation}
P_{X}^{+}(t) = \frac{\sum_{\{(i,j)|G^{\rm bin}_{X, i}(t) = 1, G^{\rm bin}_{X, j}(t) = 1\}} A_{ij}(t)}{\sum_{\{(i,j)| G^{\rm bin}_{X, j}(t) = 1\}} A_{ij}(t)} \quad.
\end{equation}

Similarly,  $P_{X}^{-}(t)$ is the conditional probability that a student is above the $X$th percentile of grades given that his or her friend is below the $X$th percentile
\begin{equation}
P_{X}^{-}(t) = \frac{\sum_{\{(i,j)|G^{\rm bin}_{X, i}(t) = 1, G^{\rm bin}_{X, j}(t) = 0\}} A_{ij}(t)}{\sum_{\{(i,j)| G^{\rm bin}_{X, j}(t) = 0\}} A_{ij}(t)} \quad.
\end{equation}
We then define the {\em conditional increase} in probability, $I_{X}(t)$ as the fraction,  

\begin{equation}
I_{X}(t) = \frac{P_{X}^{+}(t) - P_{X}^{-}(t)}{P_{X}^{-}(t)} \times 100\%
\end{equation}

\item
For scalar variables one standard way to measure homophily is the Pearson correlation coefficient $r$ across all friendship pairs which is sometimes called {\em  assortativity coefficient} \cite{newman2010homophily}.

\begin{equation}\label{Eq:rho_def}
r = \frac{\sum_{ij} (A_{ij} - k_i k_j / M)x_i x_j}{\sum_{ij} (k_i \delta_{ij} - k_i k_j / M) x_i x_j} \quad , 
\end{equation}

where $\delta_{ij}$ is a Kroneker delta,  and $M = \sum_{ij}A_{ij}$. 
\end{enumerate}

\subsubsection*{GPA distance}
We define the average {\em GPA distance} between friends for a given vector of their GPA, $G_i(t)$, and a given fixed friendship network $F_{ij}$ as 
\begin{equation}
       D_F(t) = \langle | G_i(t) - G_j(t) |  \rangle_{\{(i,j) | F_{ij} = 1\}} \quad,
\end{equation}
where $\langle . \rangle_{\{i,j\}}$ means average over all pairs of $i$ and $j$ satisfying the condition. 
We then  consider several groups of friends. 

{\em Discontinued friends} are students that are friends at time $t = 1$, but are not friends at time $t = T$. $F^{\rm disc}_{ij} = 1$ if $A_{ij}(1) = 1$, and $A_{ij}(T) = 0$. 

{\em New friends} are students that are friends at time $t = T$ but are not friends at time $t = 1$. $F^{\rm new}_{ij} = 1$ if $A_{ij}(1) = 0$ and $A_{ij}(T) = 1$. 

{\em Stable friends} are students that were friends both at time $t = 1$ and $t = T$.  $F^{\rm st}_{ij} = 1$ if $A_{ij}(1) = 1$ and $A_{ij}(T) = 1$. 

Then we can compare GPA distances between discontinued friends $D^{\rm disc}(t) = D_{F^{\rm disc}}(t)$, new friends $D^{\rm new}(t) = D_{F^{\rm new}}(t)$, and stable friends $D^{\rm st}(t) = D_{F^{\rm st}}(t)$. In the same manner we define the distance for a fixed GPA, 
\begin{equation}
        D_F = \langle | \bar{G}_i - \bar{G}_j |  \rangle_{\{(i,j) | F_{ij} = 1\}} \quad. 
\end{equation}
To compare GPA distances between new $D^{\rm new}$ and discontinued $D^{\rm disc}$ friends we perform a two sample Students' t-test.

\subsubsection*{Regression model}
To understand the influence of different covariates on students' performance, we use a simple regression 
model (\ref{Eq:regression}). We assume that students' GPAs at time $t$ depends on their GPA at the previous 
time step $t-1$, on the average GPA of their friends at the previous time step $t-1$, on  their gender and the 
years of studies\footnote{Some authors suggest to include GPA of friends for several time lags ($t$ and $t-1$ or $t-1$ and $t-2$) 
to be able to disentangle network peer influence from social selection. However, it was argued that this approach 
alone can not disentangle the two mechanisms \cite{shalizi2011homophily}.}, 

\begin{multline}\label{Eq:regression}
G_{i}(t) = \alpha_1 G_{i}(t-1) + \frac{\alpha_2}{k_i(t)} \sum_{j=1}^N A_{ij}G_{j}(t-1) + \\ \sum_{\kappa}\beta_{\kappa}{Y_{\kappa,i}} + \gamma S_i + c + \epsilon_t \quad,
\end{multline}
where $k_i(t)$ is the degree (number of friends) of student $i$ at $t$, and $\epsilon_t$ denotes white noise. 
$S_i$ is a binary variable representing the gender of a student (1 corresponds to male) and $Y_{\kappa,i}$ 
are set of binary variables, $Y_{\kappa,i} = 1$ if student $i$ is currently at the $\kappa$ year of its studies, 
and $Y_{\kappa,i} = 0$, otherwise.

To account for multiple observations of the same individual we use generalized 
estimating equations (GEE)  \cite{liang1986longitudinal}. A GEE is 
used to estimate the parameters of a generalized linear model with a possible unknown correlation between outcomes. 
To compute $p$-values for the parameters estimates we use a Wald test \cite{wald1943tests}. 


\begin{figure}
\begin{center}
\includegraphics[width=8cm]{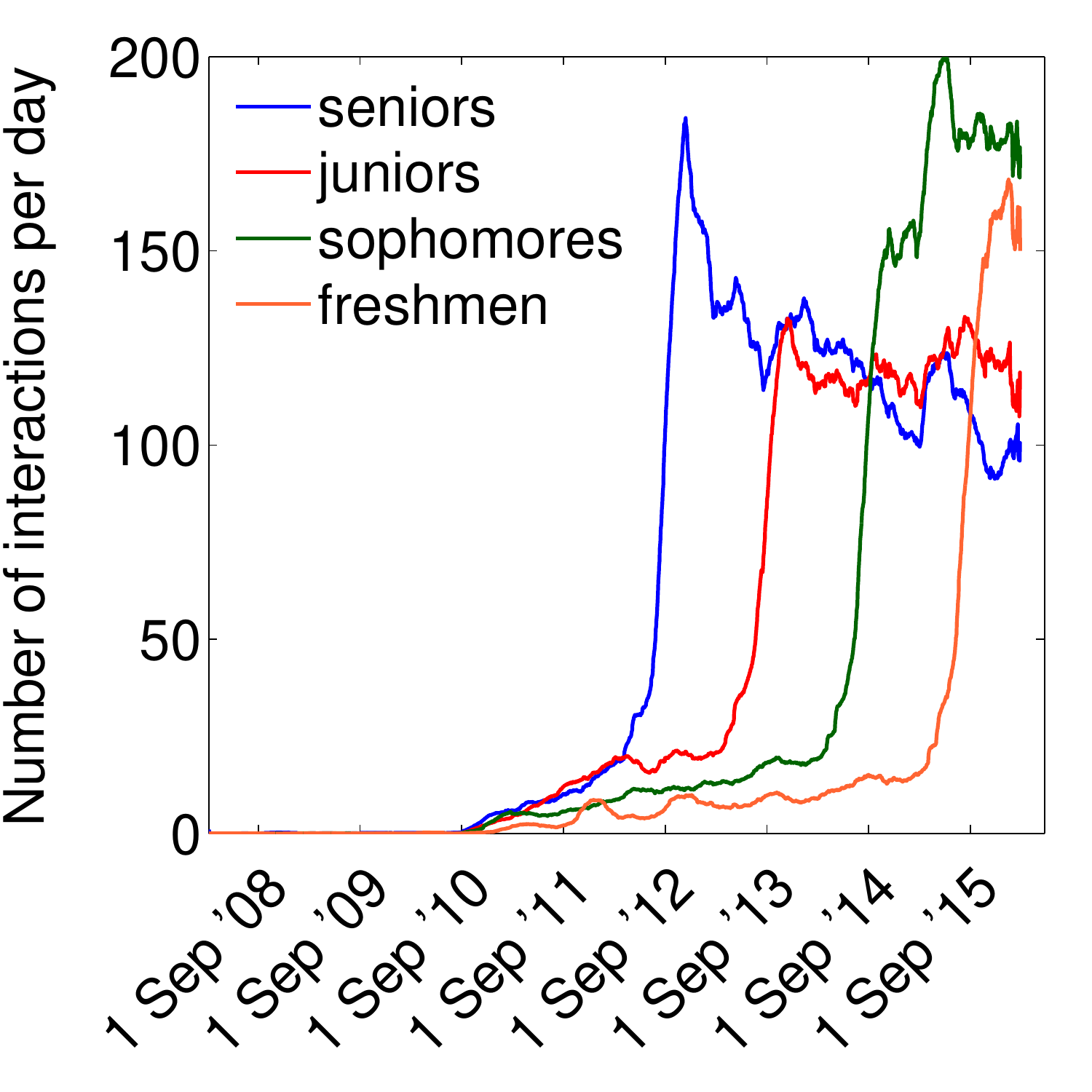}
\caption{The average number of interaction (''likes``) per day between university students is presented for each cohort. The maximum observed value is $200$ or 0.13 ``likes'' per day per student. The steep increase in September marks the beginning of studies. Some students knew each other before the matriculation.
}
\label{Fig:history}
\end{center}
\end{figure}

\begin{figure}
\begin{center}
\includegraphics[width= 8cm]{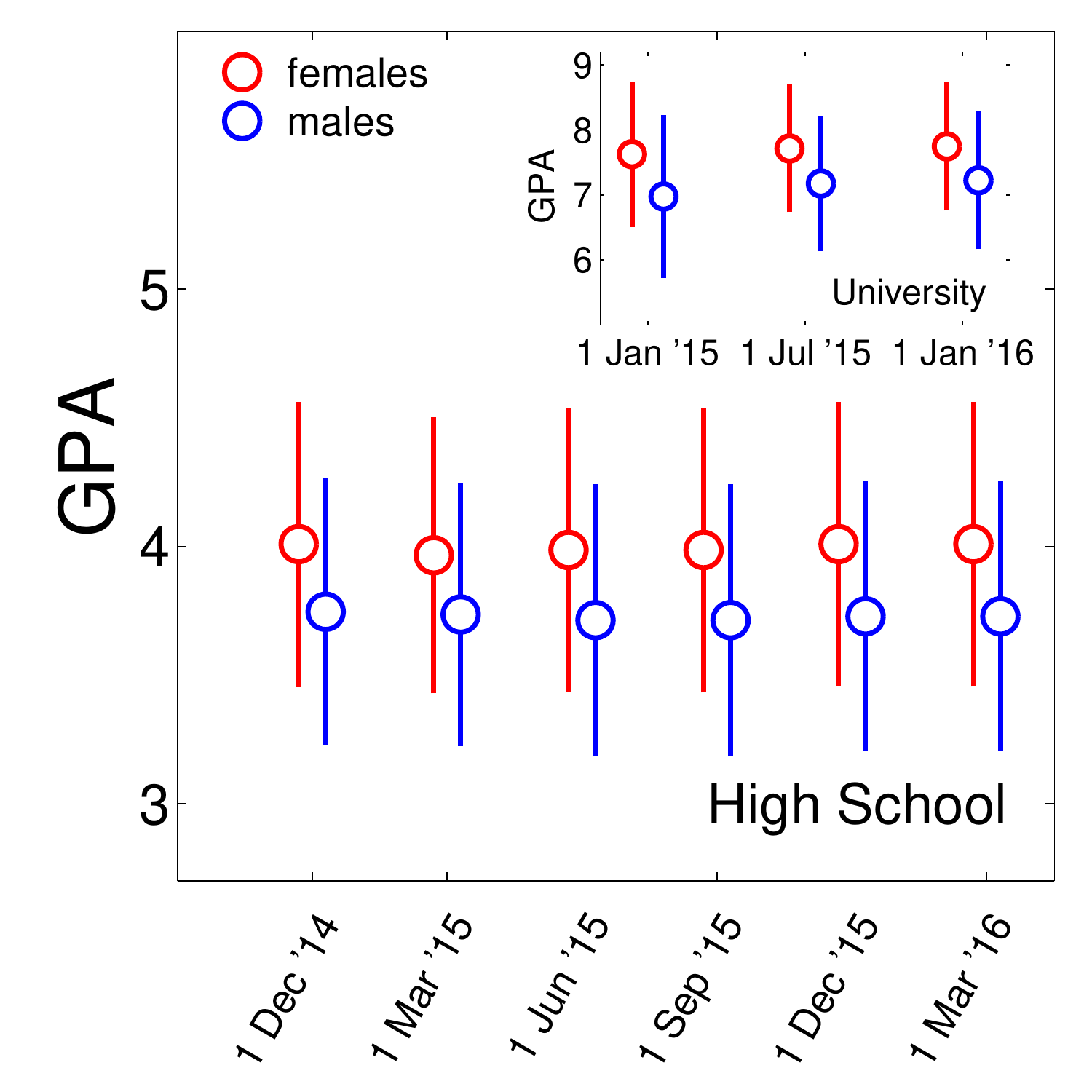}
\caption{Average GPA for high school and university students (inset) over time. Results are shown as mean values $\pm$ standard deviations. 
Females have better grades on average. GPAs and their variance do practically not change with time.}
\label{Fig:grade_diff}
\end{center}
\end{figure}

\begin{figure}
\begin{center}
\includegraphics[width= 8cm]{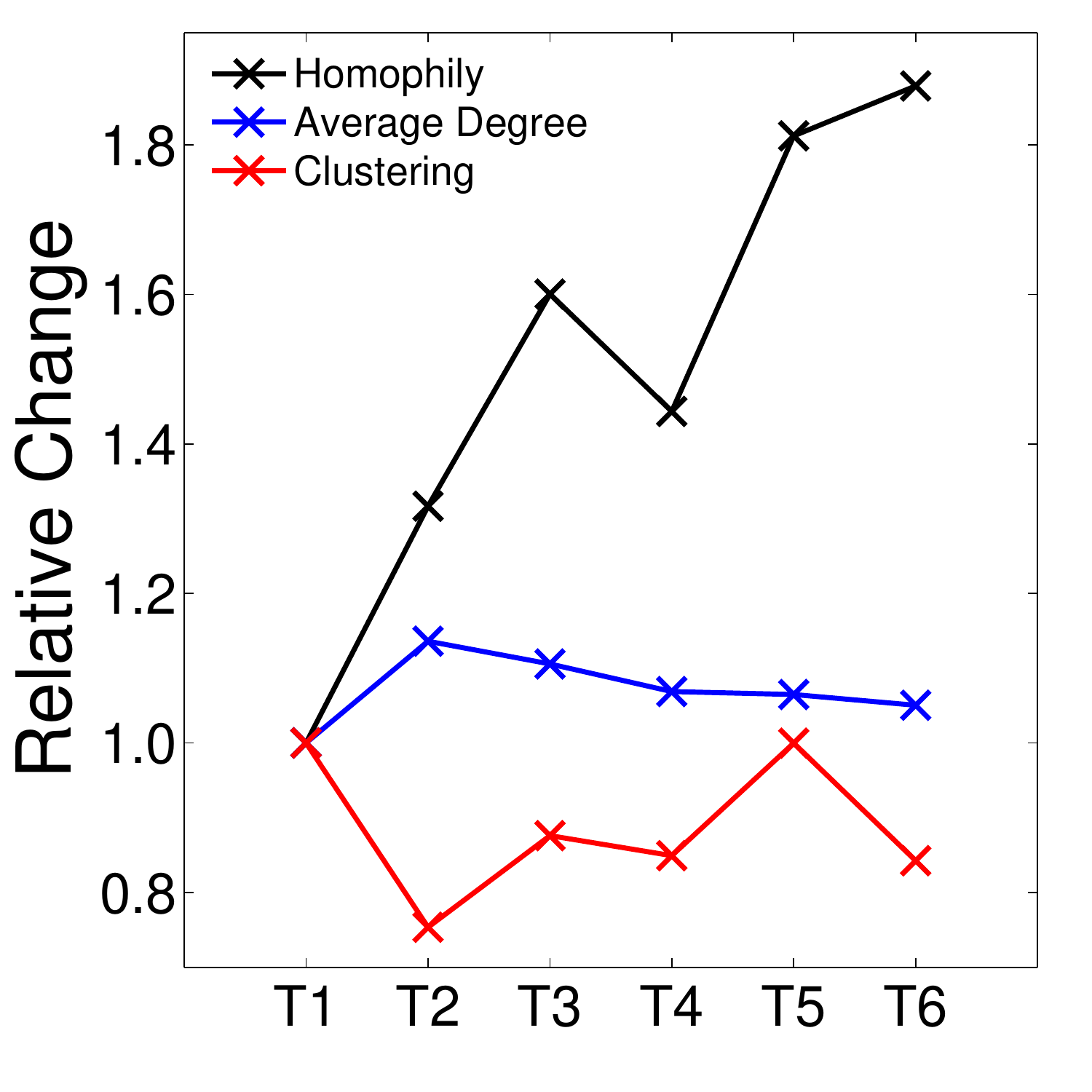}
\caption{In the high school data the network is getting more connected over time. It is therefore possible to re-define new time intervals in such a way that 
for each time interval the average degree in the network is approximately the same. Clearly the homophily index $H$ increases as before, 
indicating that the degree is not an explanatory variable. The same argument holds for the clustering coefficient.}
\label{Fig:network_adj}
\end{center}
\end{figure}

\begin{figure}
\begin{center}
\includegraphics[width= 8cm]{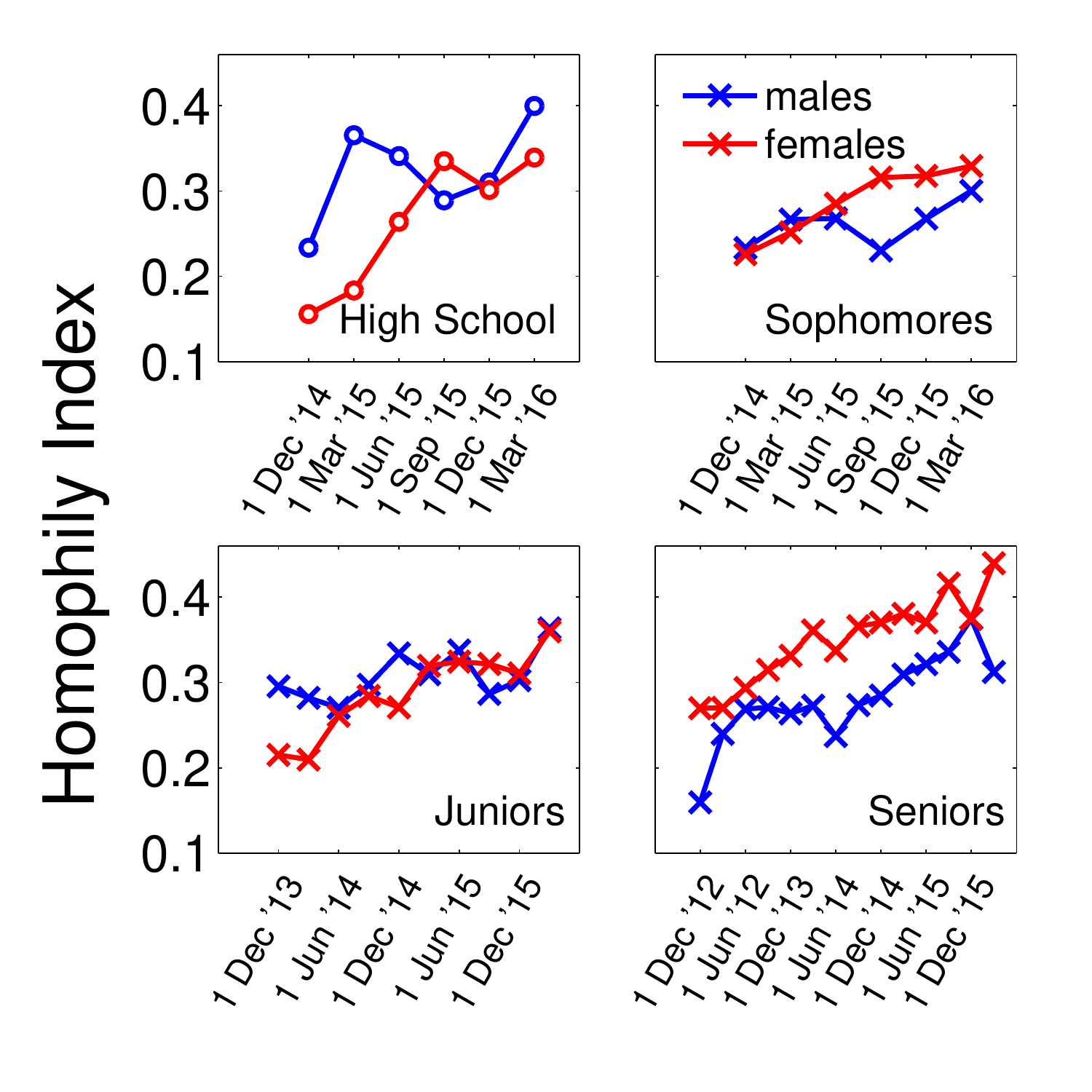} 
\caption{There are no consistent differences in gender. While both genders show about the same increase over time, 
it is larger for females in the sophomore and senior groups, and larger for males for the high school students and juniors.}
\label{Fig:gender}
\end{center}
\end{figure}

\begin{figure}
\begin{center}
\includegraphics[width= 8cm]{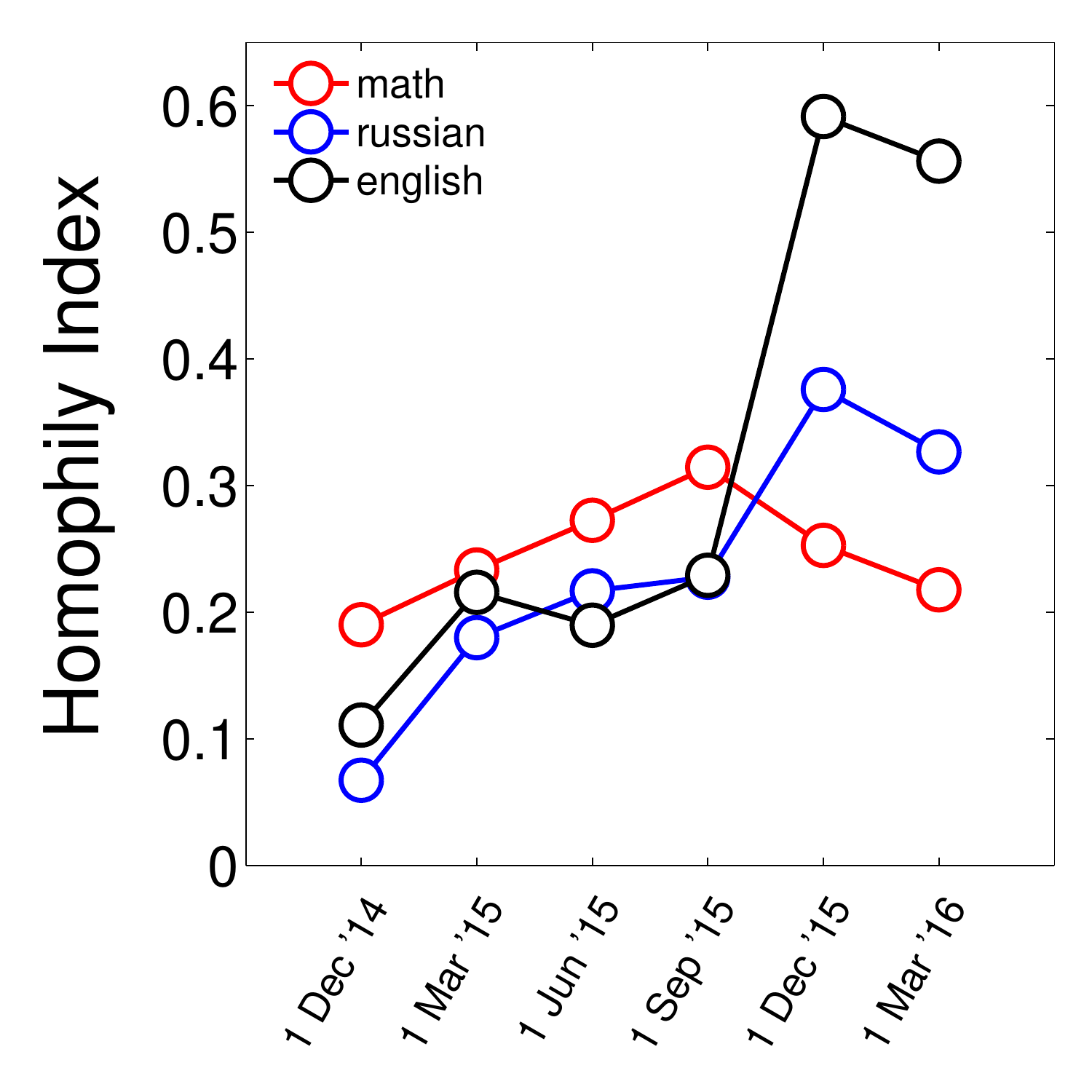}
\caption{Homophily increase  varies from subject to subject. Since there are only 4 possible values of grades (scores) possible 
for the individual subjects, we expect to observe less stable results than for the GPA. However, the general pattern of homophily increase 
over time holds, for mathematics it is not much pronounced.}
\label{Fig:subjects}
\end{center}
\end{figure}

\clearpage

\clearpage

\begin{table}[h!]
\begin{center}
\begin{tabular}{cccc}
& All students & Females & Males \\
\hline
$\langle \bar{G}^{\rm HS}_{i} \rangle_i$ & 3.85 (0.53) & 3.99 (0.53) & 3.73 (0.50) \\
$\langle \bar{G}^{\rm U, 1}_{i} \rangle_i$ & 7.32 (1.02) & 7.45 (0.97) & 7.20 (1.06) \\
$\langle \bar{G}^{\rm U, 2}_{i} \rangle_i$ & 7.01 (1.11) & 7.25 (1.05) & 6.71 (1.11) \\
$\langle \bar{G}^{\rm U, 3}_{i} \rangle_i$ & 7.40 (1.26) & 7.69 (1.15) & 7.09 (1.29) \\
$\langle \bar{G}^{\rm U, 4}_{i} \rangle_i$ & 7.35 (1.21) & 7.68 (1.10) & 6.98 (1.23) \\
\hline
\end{tabular}
\end{center}
\caption{Descriptive statistics of students' GPA scores across the whole period of their studies. $\langle . \rangle_i$ means average over all students in the group. Mean values and standard deviations (in brackets) are presented. Females have better grades than males on average.}
\label{Tbl:gpa_statistics}
\end{table}

\begin{table}[h!]
\begin{center}
\begin{tabular}{llrrr}
&& $D^{\rm disc}$ & $D^{\rm new}$ & $p$-value \\  
\hline
High School & Data & 0.56 & 0.53 & 0.02 \\
Model& $\theta = 0.55$ & 0.57 &  0.52 & 0.004 \\
\hline
Sophomores & Data & 1.17  & 1.08 & $< 10^{-8}$ \\
Model&  $\theta = 0.60$ & 1.20 & 1.10 & $< 10^{-7}$ \\
\hline
Juniors & Data & 1.22 & 1.14 & $< 10^{-5}$ \\
Model&  $\theta = 0.61$ & 1.25 & 1.17 & $< 10^{-5}$ \\
\hline
Seniors & Data & 1.22 & 1.11 & $< 10^{-8}$ \\
Model&  $\theta = 0.58$ & 1.25 & 1.13 & $< 10^{-7}$ \\
\hline
\end{tabular}
\end{center}
\caption{Re-organization of the students' network over time. The GPA distance for new friends is consistently and significantly smaller (tested with two-sample Students' test) than the GPA distance for discontinued friends in the observed data. 
Comparable results are obtained with the model.}
\label{Tbl:reorganization}    
\end{table}

\begin{table}[h!]
\begin{center}
\begin{tabular}{lrrrr}
& Parameter & Estimate & Std.Err & p-value \\
\hline
High School &$\alpha_1^{*****}$ & 0.905 &  0.013 &  $< 10^{-15}$ \\
&$\alpha_2$ & -0.003 & 0.006 &  0.56 \\
&$\gamma^{***}$ & -0.030 & 0.009 & $< 10^{-3}$ \\
&$\beta$ (5th year) & 0.020 & 0.015 &  0.18 \\
&6th year & -0.022 & 0.016 & 0.17 \\
&7th year & 0.012 & 0.014 & 0.37 \\
&8th year & -0.025 & 0.013 & 0.06 \\
&9th year & -0.026 & 0.013 & 0.05 \\
&10th year & -0.030 & 0.023 &  0.19 \\ 
&$c^{*****}$ & 0.401 & 0.061 & $< 10^{-10}$ \\
\hline
University&$\alpha_1^{*****}$ & 0.879 & 0.009 & $< 10^{-15}$ \\
&$\alpha_2$ & 0.005 & 0.012 & 0.67 \\
&$\gamma^{***} $ & -0.103 & 0.018 & $< 10^{-7}$ \\
&$\beta$ (3rd year) & 0.144 & 0.025 & $< 10^{-7}$ \\
&4th year & 0.022 &  0.027 &  0.65 \\
&$c^{*****}$ & 0.769 & 0.100 & $< 10^{-13}$ \\
\hline
\end{tabular}
\end{center}
\caption{Coefficients from the regression model (Eq. \ref{Eq:regression}). The GPA at the current time point is almost fully explained by the GPA at the previous time point. The influence of gender is also significant, males have lower grades also after controlling for their previous GPA. The average GPA of friends at the previous time point is not significant.}
\label{Tbl:regression}
\end{table}

\end{document}